\documentstyle[preprint,psfig,aps]{revtex}
\newcommand{\lan}{\langle}
\newcommand{\ran}{\rangle}

\begin{document}
\title{Memory effects in response functions of driven vortex matter} 
\author{Henrik Jeldtoft Jensen and Mario Nicodemi}
\pagestyle{myheadings}
\address{Department of Mathematics, Imperial College, 180 Queen's Gate, 
London SW7 2BZ, UK}

\maketitle
\date{\today}
\begin{abstract}
Vortex flow in driven type II superconductors shows strong 
memory and history dependent effects. Here, we study a schematic 
microscopic model of driven vortices to propose a scenario for a broad set 
of these kind of phenomena ranging from {\em ``rejuvenation''} and 
{\em ``stiffening''} of the system response, to {\em ``memory''} 
and {\em ``irreversibility''} in I-V characteristics. 
\end{abstract}
\vskip1pc


An important discovery in the dynamics of vortices driven by an external 
current in type II superconductors is the presence of strong ``memory'' and 
history dependent effects in vortex flow (see for instance 
\cite{revs,tsuei,kwok,higgins,andrei,paltiel} and ref.s therein). 
These {\em off-equilibrium} phenomena, particularly important at low 
temperatures, crucially affect the overall system behaviour and, thus, have 
also important consequences for technological applications. 
Here we present a microscopic model to explain their origin and their 
connections with other ``aging'' phenomena of vortex matter such as the rearrangement of vortex domains during creep. 
In relation to recent experimental results \cite{higgins,andrei}, we discuss 
in particular the nature of ``memory'' effects observed in the response of 
the system to an external drive, i.e., the I-V characteristic.
Our model explains the peculiar form of ``memory'' observed in vortex flow 
at finite $T$, which we call an {\em ``imperfect memory''}.
In the same framework one can explain other ``anomalous'' properties such as 
those of time dependent critical currents. 
The essential step is to identify the relevant time scales in the dynamics. 
Finally, we stress the similarities and the important differences with 
other ``glassy'' systems, such as supercooled liquids 
or disordered magnets (see \cite{glass_rev,bouchaud} and ref.s therein). 

We consider a schematic coarse grained model introduced to describe anomalous 
relaxation of vortices in superconductors, namely a system of repulsive 
particles wandering in a pinning landscape in presence of an external drive. 
The model describes several phenomena of vortex 
physics, ranging from a reentrant phase diagram in the $(B,T)$ plane, 
to the ``anomalous second peak'' in magnetisation loops 
(the ``fishtail'' effects), glassiness  and ``aging'' 
of slow magnetic relaxation, the ``anomalous creep'' at very 
low temperatures, and many others \cite{nj_1}. 


A system of straight parallel vortex lines, corresponding to a magnetic field 
$B$ along the $z$-axis, interacts via a potential \cite{revs}: 
$U(r)= {\phi^2_0\over 2\pi\lambda'^2}
\left[ K_0(r/\lambda')- K_0(r/\xi')\right]$, 
$K_0$ being the MacDonald function, $\xi$ and $\lambda$ the correlation 
and penetration lengths 
($\xi'=c\xi/\sqrt{2}$, $\lambda ' = c\lambda$, $c=(1-B/B_{c2})^{-1/2}$). 
The typical high vortex densities and long $\lambda$ imply that 
the vortex system is strongly interacting. 
To make it theoretically more tractable, 
as proposed in \cite{bassler,nj_1}, one can coarse grain in the $xy$-plane 
by introducing a square grid of lattice spacing $l$ 
of the order of the London length, $\lambda$. 
The number of vortices on the $i$-th coarse grained cell, $n_i$, 
is an integer number smaller than $N_{c2}=B_{c2}l^2/\phi_0$ \cite{nj_1} 
($B_{c2}$ is the upper critical field and $\phi_0=hc/2e$ is the flux quantum). 
The coarse grained interaction Hamiltonian is thus \cite{nj_1}:
${\cal H}= \frac{1}{2} \sum_{ij} n_i A_{ij} n_j 
-\frac{1}{2} \sum_i A_{ii} n_i - \sum_i A^p_i n_i$.
The first two terms describe the repulsion between the vortices and their 
self energy, and the last the interaction with a random pinning background. 
For sake of simplicity, we consider the simplest version of ${\cal H}$: 
we choose $A_{ii} = A_0=1$; $A_{ij}=A_1<A_0$ if $i$ and $j$ are nearest 
neighbours; $A_{ij}=0$ otherwise; the random pinning is delta-distributed 
$P(A^p)=(1-p)\delta(A^p)+p\delta(A^p-A^p_0)$.

In analogy with computer investigation of dynamical processes in 
fluids, the time evolution of the model is simulated by a Monte Carlo 
Kawasaki dynamics \cite{Binder} 
on a square lattice of size $L$ \cite{nota_L} at a temperature $T$. 
The system is periodic in the $y$-direction. The two edges parallel to the 
$y$-axis are in contact with a vortex reservoir, i.e., an 
external magnetic field, of density $N_{ext}$. 
Particles can enter and leave the system only through the reservoir. 
The above model, called ROM, is described in full details in \cite{nj_1}. 

The system is prepared by zero field cooling and then increasing $N_{ext}$ 
at constant rate up to the working value (here, $N_{ext}=10$). 
Then we monitor the system relaxation after applying a 
drive, $I$ (due to the Lorentz force), in the $y$-direction. 
As in similar driven lattice gases \cite{KLS}, the effect of the drive 
is simulated by introducing a bias in the Metropolis coupling of the 
system to the thermal bath: a particle can jump to a neighbouring 
site with a probability min$\{1,\exp[-(\Delta{\cal H}-\epsilon I)/T]\}$. 
Here, $\Delta{\cal H}$ is the change in ${\cal H}$ after the jump 
and $\epsilon = \pm 1$ for a particle trying to hop along or opposite 
to the direction of the drive and $\epsilon = 0$ if orthogonal jumps occur. 
A drive $I$ generates a voltage $V$ \cite{hyman}:
\begin{equation}
V(t)=\lan v_a(t) \ran
\end{equation}
where $v_a(t)={1\over 2\Delta t}\int_{t-\Delta t}^{t+\Delta t} v(t')dt'$ 
is an average vortex ``velocity'' in a small interval around the time $t$. 
We consider such an average to improve the statistics on $V(t)$ 
and choose $\Delta t$ accordingly. 
Here, $t$ is the Monte Carlo time (measured in units of single attempted 
update per degree of freedom), 
$v(t)= {1\over L} \sum_i v_i(t)$ is the instantaneous flow ``velocity'', 
$v_i(t)=\pm 1$ if the vortex $i$ at time $t$ moves along or opposite
to the direction of the drive $I$ and $v_i=0$ otherwise. 
The data presented below are averaged over up to 3072 realizations 
of the pinning background.

\begin{figure}[ht]
\vspace{-1.6cm}
\centerline{\hspace{-3cm}
\psfig{figure=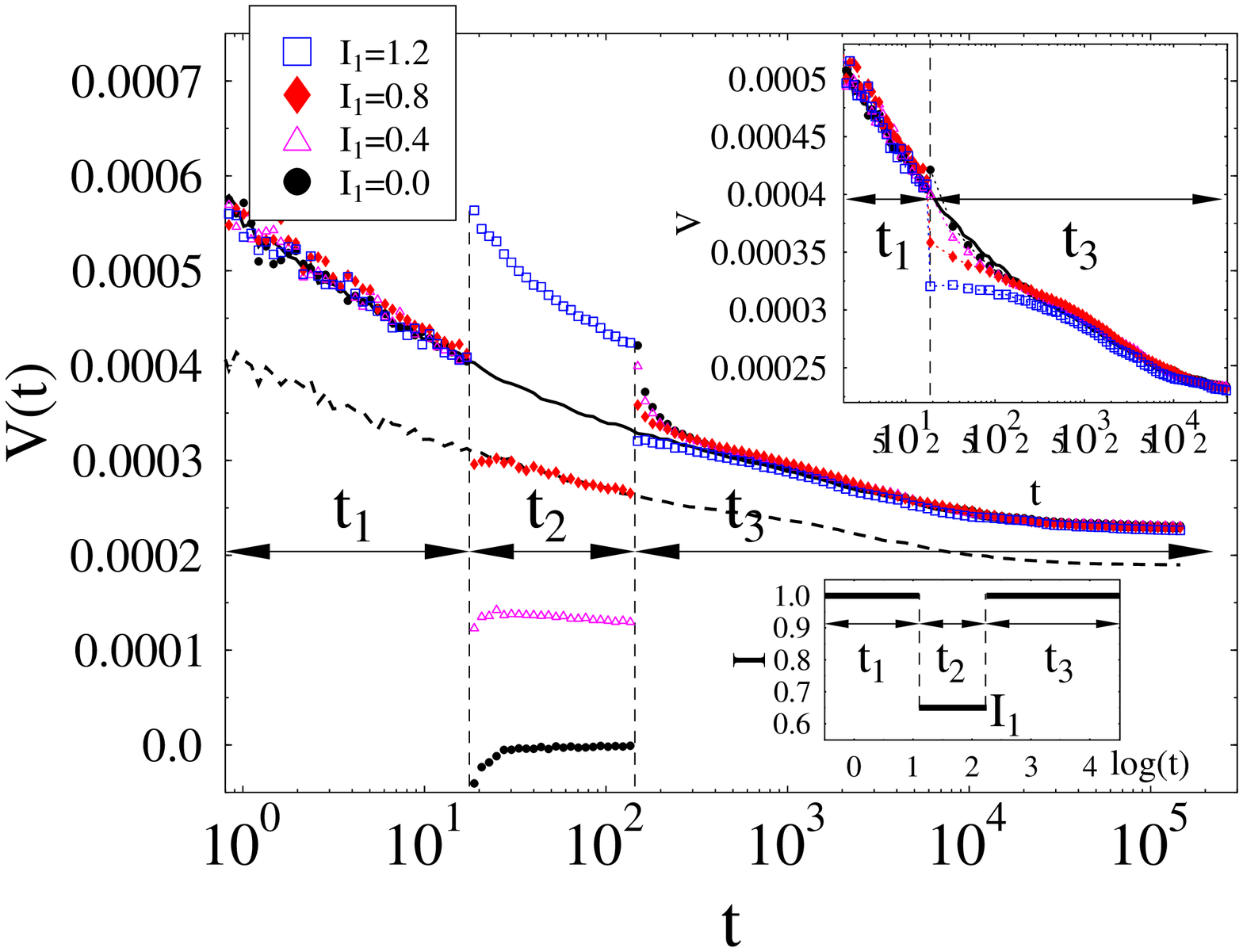,width=10.5cm,angle=-90}}
\vspace{-1.9cm}
\caption{At $T=0.1$ for a drive $I=1$, the voltage, $V(t)$, is plotted 
as a function of time. As shown in the lower inset, 
after a time lag $t_1$, the drive is abruptly 
changed to $I_1$ for a time $t_2$ and finally it is set back to its 
previous value. When $I$ is switched to $I_1$ the system seems to 
{\em ``rejuvenate''}: it suddenly restarts its relaxation along the path 
it would have had if $I=I_1$ at all times (consider the 
continuous and dashed bold curves, corresponding to $I=I_1=1$ and 
$I=I_1=0.8$, plotted for comparison). 
By restoring $I$ after $t_2$, the system shows a strong form of ``memory'': 
if $t_2$ and $I_1$ are small enough (see text) the relaxation of 
$V(t)$ restarts where it was at $t_1$. However, if $t_2$ and $I_1$ are 
too large, this is not the case, as shown in the upper inset. 
In this sense, the above is an {\em ``imperfect memory''}.} 
\label{f1}
\end{figure}

To characterise the properties of history dependent effects, we consider 
a striking manifestation of ``memory'' observed in experiments where the
drive is cyclically changed \cite{andrei}. A drive $I$ is applied to the system 
and, after a time $t_1$, abruptly changed to a new value $I_1$; finally, 
after waiting a time $t_2$, the previous $I$ is restored and the system 
evolves for a further $t_3$ (see lower inset of Fig.\ref{f1}). 
The measured $V(t)$ is shown in the main panel of Fig.\ref{f1} for $T=0.1$. 
A first observation is that after the switch to $I_1$ the system 
seems to abruptly reinitiate its relaxation approximately as if it has 
always been at $I_1$ (see for example the dashed curve in Fig.\ref{f1}), 
a phenomenon known as {\em ``rejuvenation''} 
in thermal cycling of spin-glasses and other glassy systems 
\cite{bouchaud}. The more surprising fact is, however, that for $I_1$ small 
enough (say $I_1\ll I^*$, $I^*$ to be quantitatively defined below) 
when the value $I$ of the drive is restored the voltage 
relaxation seems to restart from where it was at $t_1$, i.e., where it stopped 
before the application of $I_1$ (see Fig.\ref{f1}). Actually, if one ``cuts'' 
the evolution during $t_2$ and ``glues'' together those during $t_1$ and 
$t_3$, an {\em almost} perfect matching is observed (see upper inset of 
Fig.\ref{f1}). What is happening during $t_2$ is that the system is trapped 
in some metastable states, but {\em not} completely frozen as shown by a small 
magnetic, as well as voltage, relaxation. 
These non trivial ``memory'' effects are experimentally found in vortex matter 
\cite{andrei} and glassy systems \cite{bouchaud}. We call them 
a form of {\em ``imperfect memory''}, because they tend to disappear when the 
time spent at $I_1$ becomes too long or, equivalently 
(as explained below) when, for a given $t_2$, $I_1$ becomes too high, 
as clearly shown in the inset of Fig.\ref{f1}. 
These ``rejuvenation and memory'' effects are very general in
 off-equilibrium ``glassy'' dynamics and can be found well above the ideal 
glass transition point, $T_c$ \cite{nj_1}, 
whenever the observation time scales are 
shorter than the intrinsic relaxation scale $\tau_V$ (to be defined below). 

\begin{figure}[ht]
\vspace{-1.6cm}
\centerline{\hspace{-3cm}
\psfig{figure=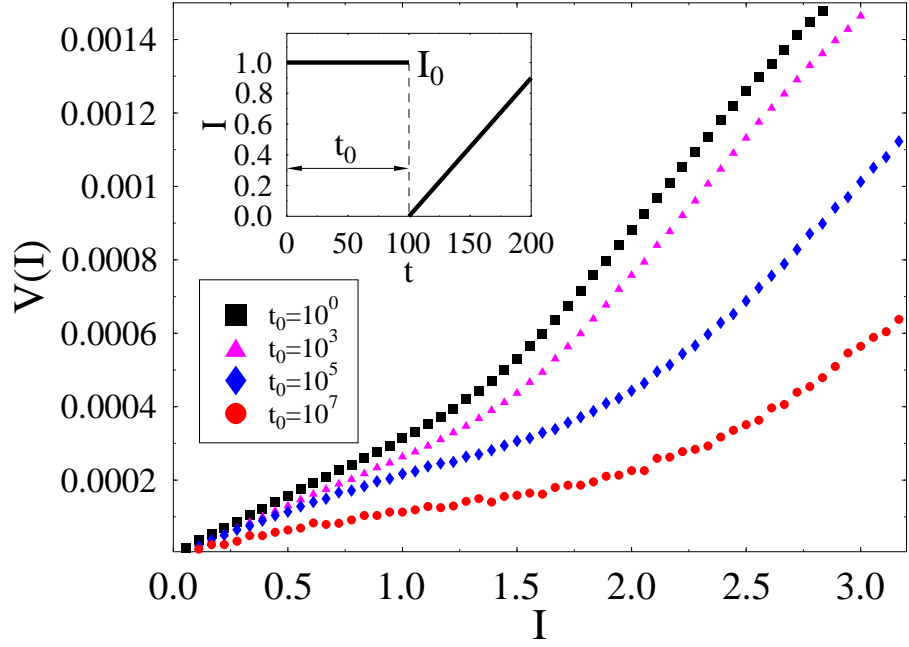,width=10.5cm,angle=-90}}
\vspace{-1.9cm}
\caption{The I-V obtained at $T=0.1$ by ramping $I$ after keeping the system 
in presence of a drive $I_0=1$ for a time $t_0$ as shown in the inset. 
The response, $V$, is ``aging'' (i.e., depends on $t_0$) and, 
more specifically, {\em stiffening}: it is smaller the longer $t_0$.} 
\label{f2}
\end{figure}

We now turn to some additional aspects concerning the time 
dependent properties of the current-voltage characteristic. 
As in real experiments on vortex matter \cite{andrei}, 
we let the system undergo a current step of hight $I_0$ for a time $t_0$ 
before starting to record the I-V by ramping $I$, as sketched in the inset of 
Fig.\ref{f2}. Fig.\ref{f2} shows (for $T=0.1$) that the I-V depends on the 
waiting time $t_0$. The system response is ``aging'': the longer 
$t_0$ the smaller the response, a phenomenon known as {\em ``stiffening''} 
in glass formers \cite{glass_rev,bouchaud}. 
These effects are manifested in a violation of time translation invariance of 
the two times correlation functions, which have dynamical scaling properties 
\cite{nj_1} analogous to those of glass formers \cite{glass_rev}. 

These simulations also reproduce the experimentally found time dependence of 
the 
critical current \cite{andrei}. Usually, one defines an effective critical 
current, $I_c^{eff}$, as the point where $V$ becomes larger than a given 
threshold (say $V_{thr}=10^{-5}$ in our case): one then finds that 
$I_c^{eff}$ is $t_0$ and $I_0$ dependent (like in experiments \cite{andrei} 
$I_c^{eff}$ is slowly increasing with $t_0$, see Fig.\ref{f2}).

\begin{figure}[ht]
\vspace{-1.9cm}
\centerline{\hspace{-3cm}
\psfig{figure=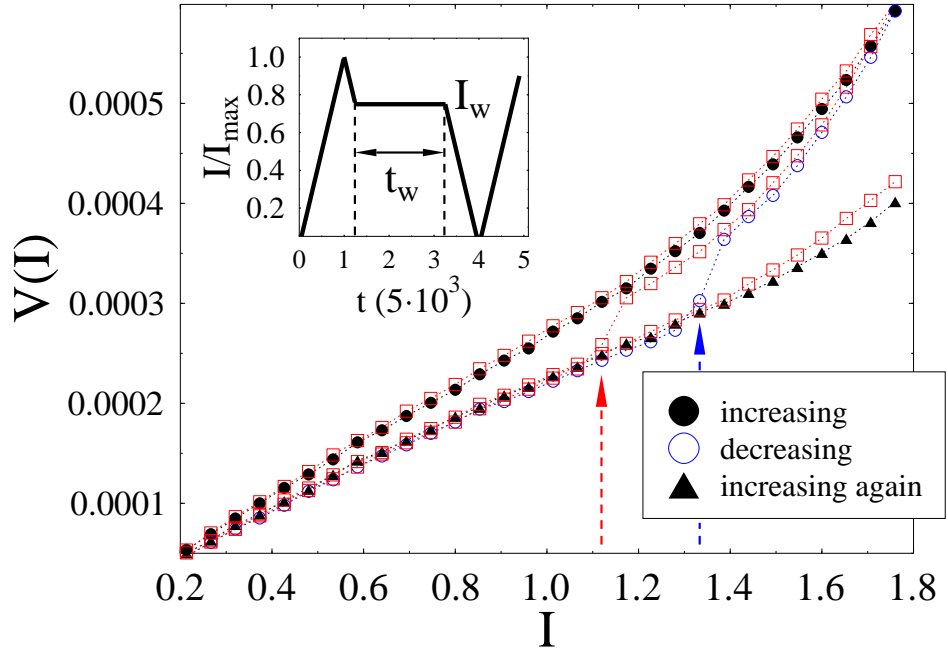,width=10.5cm,angle=-90}}
\vspace{-1.8cm}
\caption{The I-V is measured at $T=0.1$ during cycles of $I$ 
(see also the inset): 
$I$ is at first increased up to $I_{max}$ (filled circles); along the 
descending branch of the cycle (empty circles), when $I=I_w$ (in the main 
panel the $I_w$'s, for two cycles, are located by the arrows) the drive is 
kept fixed for a time 
$t_w=10^4$ and then the cycle restarted; finally, $I$ is ramped up again 
(filled triangles). 
For $I>I_w$, the first increasing ramp and the decreasing one (resp. filled 
and empty circles) do not completely match, showing {\em irreversibility} in 
the I-V. After waiting $t_w$ at $I_w$, a much larger separation is seen. 
However, by raising $I$ again (filled triangles) a strong {\em memory} 
is observed: the system doesn't follow the first branch (filled circles), 
but the decreasing one (empty circles). 
Furthermore, in a cycle with a lower $I_w$ (squares), 
the {\em same} branches are found.} 
\label{f3}
\end{figure}

It is interesting to consider another current cycling experiment which 
outlines the concurrent presence of irreversibility and memory effects. 
The I-V is measured by ramping $I$ up to some value $I_{max}$. Then $I$ 
is ramped back to zero, but at a given value $I_{w}$ the system is let to 
evolve 
for a long time $t_w$. Finally, $I$ is ramped up again (see inset of 
Fig.\ref{f3}). The resulting {\em irreversible} $V(I)$ 
is shown in Fig.\ref{f3}. 
For $I>I_{w}$ the decreasing branch of the plot (empty circles) 
slightly deviates from the increasing one (filled circles), showing the 
appearance of {\em irreversibility}. This is even more apparent after $t_w$: 
for $I<I_{w}$ the two paths are clearly different. Interestingly, upon 
increasing $I$ again (filled triangles), 
$V(I)$ doesn't match the first increasing branch, but the latest, the 
decreasing one: in this sense there is 
coexistence of memory and irreversibility.
Also very interesting is that by repeating the cycle with a new $I_{w}$ 
(squares), the system approximately follows the {\em same} branches. 
This non-reversible behaviour is also found in other glassy systems 
\cite{bouchaud}. However, spin glasses, for instance, show the presence 
of the so called {\em chaos effects} \cite{bouchaud}. 
The chaos effect is absent in our system as it is also in other ordinary 
glass formers \cite{bouchaud}. 
This kind of interplay between irreversibility and memory can be checked
experimentally in superconductors and thereby assess the present scenario.  

The above experiments outline a set of very important ``history'' dependent 
phenomena (``rejuvenation'', ``imperfect memory'', ``stiffening'' and 
``irreversibility'' of response, etc...). 
A crucial theoretical step to understand them is the identification of the 
characteristic time scales of the driven dynamics. This we now discuss.

\begin{figure}[ht]
\vspace{-1.6cm}
\centerline{\hspace{-3cm}
\psfig{figure=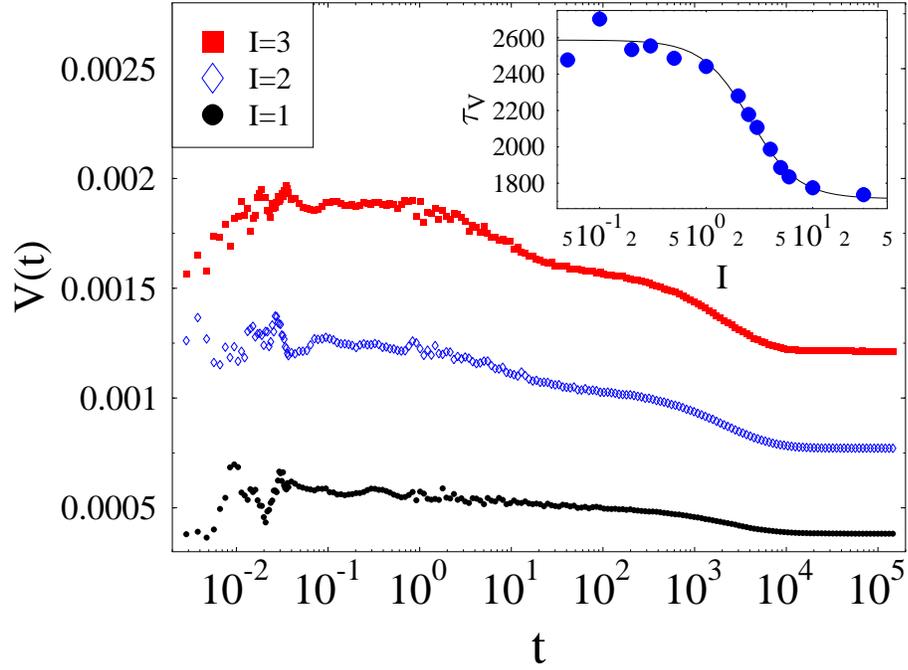,width=10.5cm,angle=-90}}
\vspace{-1.9cm}
\caption{The time evolution of the response function, $V(t)$, for the shown 
values of the drive $I$ (at $T=1$ and $N_{ext}=10$). 
In the asymptotic regime $V(t)$ is well 
fitted with: $V(t)\propto \exp[-(t/\tau_V)^{\beta}]$. 
{\bf Inset} The characteristic scale of relaxation, $\tau_V(I)$, as a function 
of $I$. For $I\rightarrow 0$, $\tau_V(I)$ seems to 
saturate to a {\em finite} value which implies $I_c=0$.} 
\label{f4}
\end{figure}

Upon applying to the system a small drive, $I$, its response, $V$, relaxes 
following a pattern with two very different parts: 
at first a rapidly changing non-linear response is seen, 
later followed by a very slow decrease towards stationarity 
(see $V(t)$ in Fig.\ref{f4} for $T=1$ and $I\in\{1,2,3\}$). 
For instance, for $I=3$ in a time interval $\Delta t\simeq 2\cdot 10^{-1}$, 
$V$ leaps from about zero to $\Delta V_i\sim 2\cdot 10^{-3}$, corresponding 
to a rate $r_i=\Delta V_i/\Delta t\sim 10^{-2}$. This is to be compared with 
the rate of the subsequent slow relaxation from, say, $t=2\cdot 10^{-1}$ 
to $t=10^4$, $r_f \sim -10^{-7}$: 
$r_i$ and $r_f$ differ of 5 orders of magnitude.

In agreement with experimental findings \cite{revs,higgins,andrei,danna}, 
the slow relaxation of $V(t)$ has a characteristic double step 
structure, which asymptotically can be well fitted by stretched exponentials
\cite{nota_log}: $V(t)\propto \exp(-t/\tau_V)^{\beta}$. 
The above long time fit defines the characteristic asymptotic scale, $\tau_V$, 
of relaxation. The exponent $\beta$ and $\tau_V$ 
are a function of $I$, $T$ and $N_{ext}$ (see inset Fig.\ref{f4}): 
in particular $\tau_V(I)$ decreases with $I$ and seems to approach 
a {\em finite plateau} for $I<I^*$, with $I^*\simeq O(1)$. 
In this sense, the presence of a drive $I$ makes the approach
to stationarity faster and has an effect similar to an increase in $T$. 
A reasonable fit for $\tau_V(I)$ is: 
$\tau_V(I)=(\tau_V^0-\tau_V^{\infty})/[1+(I/3I^*)^a]+\tau_V^{\infty}$, 
with $\tau_V^0=2590$, $\tau_V^{\infty}=1720$, $I^*=0.86$ and $a=2$. 

The outlined properties of $\tau_V$ clearly explain the history  
dependent effects in the experiments previously considered. 
For instance, the ``imperfect memory'', discussed in Fig.\ref{f1}, is caused by
the presence of a long, but finite, scale $\tau_V$ in the problem: 
for a given $I_1$ the system seems to be 
frozen whenever observed on times scales smaller than $\tau_V(I_1)$. 
Thus, if $t_2$ is short enough ($t_2<\tau_V(I_1)$) the system preserves a 
strong ``memory'' of its state at $t_1$. 
The weakening of such a ``memory'' found for higher currents $I_1$ 
in Fig.\ref{f1}, is also a consequence of the strong decrease of 
$\tau_V(I)$ with $I$. The phenomenon of ``rejuvenation'' (see Fig.\ref{f1}) 
is, in turn, a consequence of the presence of the extremely fast first part 
of relaxation found in $V(t)$ upon applying a drive 
and of the above long term memory. 
The existence of the slow part in the $V(t)$ relaxation 
also affects the ``stiffening'' of the response in the I-V of Fig.\ref{f2}, 
which is due to the non-stationarity of the vortex flow 
on scales smaller that $\tau_V$. 
Actually, in Fig.\ref{f2}, for a given $I$ the value of $V$ on the different 
curves corresponds to the system being probed at different stages of its 
non-stationary evolution (this also outlines that the proper definition of $I_c$ 
is the {\em asymptotic} one given later on). 
Finally, in brief, 
the fact that $\tau_V(I)$ is smaller at high currents, $I$, and 
larger at small $I$ (and $T$), is responsible for the surprisingly 
concomitant effects of irreversibility and memory of Fig.\ref{f3}. 

\begin{figure}[ht]
\vspace{-2.6cm}
\centerline{\hspace{-3cm}
\psfig{figure=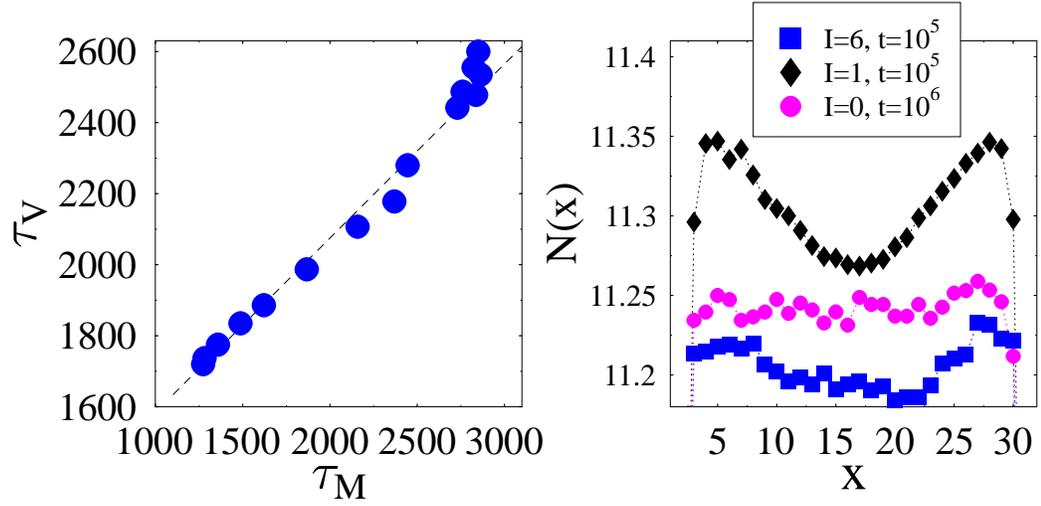,width=10.5cm,angle=-90}}
\vspace{-2.4cm}
\caption{{\bf Left} 
Vortex driven flow and creep are strictly related. Actually, 
the characteristic time scales of voltage relaxation, $\tau_V(I)$, 
and of magnetic creep, $\tau_M(I)$ (shown for $T=1$), 
are approximately proportional. 
{\bf Right } The Bean field profile, $N(x)$, present in the system during 
relaxation is shown for the plotted values of $I$ and $t$.} 
\label{f5}
\end{figure}

The origin of these time dependent properties of the driven flow, 
and in turn those of I-V's, traces back to the 
concurrent vortex creep and reorganisation of vortex domains. In fact, 
both with or without an external drive, the 
system evolves in presence of a Bean like profile (see Fig.\ref{f5}) 
which in turn relaxes 
following stretched exponentials \cite{nj_1} with a characteristic time scale 
$\tau_M$. An important discovery is that $\tau_M(I)$ and $\tau_V(I)$ are 
approximately proportional, as shown in Fig.\ref{f5}. 
This outlines that the strong non-linear, 
non-stationary voltage relaxation is structurally related to
the reorganisation of vortices during the creep (a fact 
confirmed by recent experiments \cite{paltiel}). 
By assuming that the 
system is composed of evolving vortex domains characterised by 
drive dependent (free) energy barriers which grow with the sizes of the 
domains, a hierarchy of time scales naturally arises. 
The simultaneous presence of both long time scales (``aging modes'') and 
short time scales is a simple mechanism able to describe the above 
phenomena (see \cite{bouchaud}).

Finally, let us notice that the existence at $T=1$ of a finite $\tau_V$ limit 
for $I\rightarrow 0$ (see Fig.\ref{f4}) implies that the 
flux flow can be activated for any finite $I$. This shows that the 
system's asymptotic (i.e., for $t\rightarrow\infty$) critical current, $I_c$, 
is zero. 
It is possible that the asymptotic value of $I_c$ can be
non-zero for real three dimensional systems if vortex line
cutting is negligible. The above also points out that a true divergence in 
$\tau_V(I)$, i.e., a true asymptotic finite $I_c$, can be found only for 
$T<T_c$, $T_c$ being the {\em ideal glass transition} point \cite{revs,nj_1}, 
where by definition $\tau_M(T)|_{I=0}$ diverges \cite{nj_1}.


The few key ``experiments'' we discussed in the context of our schematic 
microscopic model depict a new clear scenario for ``aging'' and 
``memory'' effects in driven vortex matter. In the present framework a broad 
range of phenomena can be understood (from response ``stiffening'' and 
``rejuvenation'', to time and history dependent critical currents, 
$I_c^{eff}$). We stress that most of these features can also be found well 
above the ideal glass transition point $T_c$. 
We have shown how creep and response functions in driven media
are related. 
Theoretically intriguing is the observation of similar behaviours found in 
vortex matter and others glass formers such as random magnets and supercooled 
liquids.

\newpage

\newpage

\newpage

\newpage

\begin{thebibliography}{40}
\bibitem{nota_L} Here
$A_1=0.28 A_0$, $A_0^p=0.3 A_0$, $N_{c2}=27$, $p=1/2$
(the ${\cal H}$ parameters can be related to the real material parameters 
\cite{nj_1}) and $L=32$. Our results are checked up to $L=128$ and robust 
to parameter changes.
\bibitem{nota_log} At lower $T$ inverse logarithmic relaxations are 
found \cite{nj_1}.
\end{thebibliography}
\end{document}